\documentclass[conference]{IEEEtran}
\IEEEoverridecommandlockouts
\usepackage{cite}
\usepackage{amsmath,amssymb,amsfonts}
\usepackage{algorithmic}
\usepackage{graphicx}
\usepackage{textcomp}
\usepackage{textgreek}
\usepackage{xcolor}
\usepackage{url}
\def\BibTeX{{\rm B\kern-.05em{\sc i\kern-.025em b}\kern-.08em
    T\kern-.1667em\lower.7ex\hbox{E}\kern-.125emX}}

\begin{document}

\title{VMAF And Variants: Towards A Unified VQA \\

\thanks{Research funded in part by grants from the US Army.}
}

\author{\IEEEauthorblockN{ Pankaj Topiwala}
\IEEEauthorblockA{\textit{FastVDO LLC} \\
Melbourne, FL 32940 \\
pankaj at fastvdo.com}
\and
\IEEEauthorblockN{ Wei Dai}
\IEEEauthorblockA{\textit{FastVDO LLC} \\
Melbourne, FL 32940 \\
daisy at fastvdo.com}
\and
\IEEEauthorblockN{ Jiangfeng Pian}
\IEEEauthorblockA{\textit{FastVDO LLC} \\
Melbourne, FL 32940 \\
pian at fastvdo.com}
\and
\IEEEauthorblockN{ Katalina Biondi}
\IEEEauthorblockA{\textit{FastVDO LLC} \\
Melbourne, FL 32940 \\
kat at fastvdo.com}
\and
\IEEEauthorblockN{ Arvind Krovvidi}
\IEEEauthorblockA{\textit{FastVDO LLC} \\
Melbourne, FL 32940 \\
arvind at fastvdo.com}
}

\maketitle

\begin{abstract}
Video quality assessment (VQA) is now a fast-growing subject,  maturing in the full reference (FR) case, yet challenging in the exploding no reference (NR) case. We investigate variants of the popular VMAF video quality assessment algorithm for the FR case, using both support vector regression and feedforward neural networks. We extend it to the  NR case, using some different features but similar learning, to develop a partially unified framework for VQA. When fully trained, FR algorithms such as VMAF perform very well on test datasets, reaching 90\%+ match in PCC and SRCC; but for predicting performance in the wild, we train/test from scratch for each database. With an 80/20 train/test split, we still achieve about 90\% performance on average in both PCC and SRCC, with up to 7-9\% gains over VMAF, using an improved motion feature and better regression.  Moreover, we even get decent performance (about 75\%) if we ignore the reference, treating FR as NR, partly justifying our attempts at unification. In the true NR case, we reduce complexity vs. leading recent algorithms VIDEVAL, RAPIQUE, yet  achieve performance within 3-5\%. Moreover, we develop a method to analyze the saliency of features, and conclude that for both VIDEVAL and RAPIQUE, a small subset of their features are providing the bulk of the performance. In short, we find encouraging improvements in trainability in FR, while  constraining training complexity against leading methods in NR, elucidating the saliency of features for feature selection.
\end{abstract}


\section{Introduction}
For us humans, vision is our most powerful sense, and the visual cortex makes up 30\% of the cerebral cortex in the brain (8\% for touch, and just 3\% for hearing) \cite{b1-30p}. And visual stimulus is typically our most informative input. Developed over eons for detecting predator (or prey) by registering movement, vision has since developed into our single all-encompassing sense. It is not surprising that as our gadgets and networks have matured in recent times, video makes up a staggering 80\%+ of all internet traffic today, a fraction that is still rising \cite{b2-csco}. That video is big business; it is highly processed and heavily monetized, by subscription, advertisement, or other means, creating a \$200B+ global market in video services. Netflix itself takes up some 37\% of network bandwidth, while YouTube serves a staggering 5B streams/day (1B hrs/day). Due to the immense bandwidth of raw video, a panoply of increasingly sophisticated compression algorithms have been developed, from H.261 to H.266, now achieving up to a staggering 1000:1 compression ratio with the latest H.266/VVC video codec \cite{b3-fv20}. Most of this video traffic is meant for human consumption, although a growing fraction is now aimed at machine processing such as machine vision (video coding for machines, VCM). And going forward, algorithm developers are looking to neural networks to supply the next-level performance (and especially for VCM). The future for video coding looks neural, first at the component level, then end-to-end. But coding is only half the problem.

Due to the volumes of video created and served globally, this industry also needs an array of objective metrics that are expected to be predictive of subjective human ratings. But for most of the past 40 years, the video coding research and development industry has been using mean-squared error-based PSNR, the most basic FR VQA. Moreover, in the encoder, an even simpler measure, the sum of absolute differences (SAD) is used instead of MSE, just to avoid multiplications! Puns aside, it is known that SAD correlates even less to subjective scores than MSE. We predict that this disconnect between the development of video coding, and its important use cases will change going forward. And from neuroscience, it is also natural that learning techniques like support vector machines and neural networks would be useful in assessing the quality of video streams by objective methods (\cite{b23-aqua} even develops a VQA for VCM). As neural methods gain a footing in VQA, methods such as architectural learning and overfitting management (such as dropouts) will be tested. For now we use the simplest methods. 

A key difference in full reference (FR) vs no reference (NR) VQA domains appears right at the source. Movie studios, broadcasters and subscription VOD services like Netflix/Amazon Prime use {\bf professional} capture and editing with high-end equipment at very high rates, creating {\bf contribution}-quality originals, while {\bf distributing} lower rate derived streams to consumers. In assessing the quality of their distribution streams, they have the full reference original to compare with. Considerable advances in FR VQA have culminated in algorithms such as VMAF from Netflix (introduced in 2016, and now updated with additional features) \cite{b4-nflx}, as well as a torrent of all-neural network methods, of which we cite just one: C3DVQA \cite{b5-c3dvqa}. These achieve 90\%+ agreement with user ratings on test databases after extensive training. C3DVQA is a complex all-NN design, with 2D CNNs to extract spatial features, and 3D ones to extract spatio-temporal ones. VMAF uses well-known fixed-function features, and a simple SVR regressor. At present, to limit the high complexity of expensive feature extraction, we prefer computationally feasible features, and apply efficient learning-based methods post feature extraction to derive methods usable in the near-term. While authors typically report only inference time, we report full training/testing time post feature extraction.

By contrast, the {\bf user-generated} content on prominent social media sites such as on YouTube, Facebook, and TikTok is typically acquired by novice users with unstable handheld smartphones (or GoPros), often in motion, and with little or no editing. These social media services lack any pristine reference to compare to, and have had to develop ad hoc methods to monitor the volumes of video emanating from their servers in a challenging no reference (NR) or blind VQA setting. This field relies on intrinsic qualities of the video to develop a measure. For this, they have in part focused on the perceived Gaussianity of natural scene statistics (NSS) and evaluating how video distortions affect or alter those statistics to create a measure of quality. An entire cottage industry has thus sprung up to create both FR and NR VQA measures, that can adequately meet the needs for stream selection and monitoring. In sum, what separates the professional FR and user-generated NR worlds is the markedly different quality of capture (in terms of sensors, stability, noise, blur, etc.). And this is also reflected in the databases we work with; see Figs. 4, 5. 

Even with the wide gulf between these domains, in this paper, we attempt a partial synthesis of some trends in FR and NR VQAs, to formulate what we call FastVDO Quality (FVQ).  It incorporates some lessons from the FR VMAF, the NSS-based assessment concepts in the NR VIIDEO \cite{b12-viid}, SLEEQ \cite{b16-sleeq}, VIDEVAL \cite{b18-videval}, and RAPIQUE \cite{b20-rapique}, and our own research over the past several years in using learning-based methods in VQA, to create one method that applies to both cases. Our contributions are as follows: (a) an improved motion feature; (b) improved SVR hyperparameters for no-search FR VQA; (c) use of neural network regression; (d) aggregation of diverse NR features and novel feature selection criterion; and (e) reduced parametric regression. 

\section{Review of VQAs and their Uses}

The aim of a VQA is to predict not individual ratings but a mean opinion score (MOS) among viewers with high correlation, as measured using correlation coefficients. For random variables X and Y, the Pearson linear Correlation Coefficient (PCC) and the Spearman Rank order Correlation Coefficient (SRCC) are given by:

\begin{equation}
\begin{aligned}
& PCC (X,Y)=(E[(X-\mu_X )(Y-\mu_Y )]/(\sigma _X \sigma_Y )),  \\ 
& SRCC=PCC(rk(X),rk(Y)),\\
& where \ rk(X)=rank \ order \ of \ X.  
\end{aligned}
\end{equation}

Note that these measures could also be calculated at the frame, or even block-level if desired (but challenging to capture user ratings); while we mainly work with video-level measures, for encoder optimization one needs deeper analysis; see the encoder optimization discussion below. PCC measures direct correlation, while SRCC only on the rank order; yet it is more directly useful in live application to stream selection. As mentioned, SAD and MSE have been the most used VQAs, despite poor correlation; see Figs. 1, 4, and \cite{b3-fv20}.

If a VQA algorithm achieves high scores for both PCC and SRCC in test databases, we envision at least three separate, increasingly larger but more demanding applications. First, the VQA can be used in {\bf stream selection} (send the best quality video), which is an elementary, typically offline application, post compression. This is perhaps the most prevalent problem faced by streamers like Netflix, Hulu and Amazon: among multiple encodings, which stream will optimize viewer appreciation. In reality, this task is further complicated by variation in instantaneous channel bandwidth, as well as transmission issues such as dropped packets, rebufferings, etc. We mainly focus on assessing the quality loss due to compression and leave aside considerations of transmission (these are generic anyway). If a VQA has a high SRCC to subjective scores, then for a given bandwidth limitation, the stream below the bandwidth limit with the highest SRCC score should be selected. A related task is video quality {\bf monitoring}: measure the predicted quality of outgoing streams. For this, both PCC and SRCC are used. Selection and monitoring are key applications.

Second, a VQA can be used in receiver video {\bf restoration} (restore for best visual quality). Such a VQA could be combined with deep learning to train blocks of video frames on the original video, which can provide effective restoration in compressed and other distorted videos \cite{b3-fv20}, \cite{b6-fv18}, \cite{b7-fv19}. This is a large and powerful application, especially when done offline. Finally, it could be used for video {\bf encoder optimization} to decide how best to encode a video with a given codec (endcode for best visual quality). While stream selection (at server) and restoration (at receiver) can require real-time performance, and thus pose complexity constraints, it is the encoding application that is by far the most challenging; we will focus first on this application. The issue is that all modern encoders rely on using rate-distortion optimization (RDO) \cite{b8-rdo} to make decisions, based on an interplay between distortion D, and the rate R, to optimize the Lagrangian (where \textlambda \ is a constant called a Lagrange multiplier):

\begin{equation}
\begin{aligned}
& L= D + \lambda R = \sum\limits_{i} D_i +\lambda R_i  ; \\
&  \delta L=0 => \delta L_i=0  \\
& \implies \lambda = - D_i/R_i , \ a \ constant.
\end{aligned}
\end{equation}

Thus, given any number of independent parameters to optimize (e.g., various pixel quantizers), these are jointly optimized when the slopes of negative distortion over rate are all equal \cite{b9-wav}. In general, the RDO is more complicated, but still essential. In coding a 4K or 8K video, a modern encoder such as VVC may make millions of RDO decisions per second, on everything from mode selection and motion estimation to quantization and filtering decisions. These are typically done at the block-level, so are computationally costly. Further, since many video applications require real-time encoding (e.g., transmission of live events in news or sports), usually done in hardware, this puts severe constraints on how RDO is actually computed. Now in rate-distortion analysis, the rate R is straightforward: how many bits it takes to encode the data (though even this is estimated to save cycles, not computed). But what to use for the distortion D, comparing a coded MxN block B to the reference version, is more open. Typically, the simple mean squared error (MSE) or L2-norm is used to represent the block-based spatial error E(k,spat). As mentioned, this is further simplified to just the Sum of Absolute Differences (SAD, or L1-norm).

\begin{equation}
\begin{aligned}
& E_{k,spat}=SAD=\sum\limits_{i,j=1}^{M,N} \lvert B_{ref,i,j} -B_{coded,i,j} \rvert \\
& = \lvert \lvert F_{ref}-F_{coded} \rvert \rvert, \ the \ L1 \ norm.
\end{aligned}
\end{equation}

\section{FVMAF: VMAF + Improved Motion}

Now if an original (uncompressed) video sequence is a set of frames F(k), k=0,…,K, the popular VMAF \cite{b4-nflx} algorithm uses two known IQAs, Visual Information Fidelity (VIF) and Detail Loss Measure (DLM), as well as the Sum of Absolute Frame Difference (SAFD) as a motion feature (Netflix calls this Mean of Co-located Pixel Difference), where the L1-norm is used. Herein, we will refer to this feature as M for motion, which is used along with 4 scale-based VIF features, and DLM (6 total).

\begin{equation}
\begin{aligned}
& SAFD = \sum\limits_{k=1}^{K} \lvert \lvert F(k) - F(k-1) \rvert \rvert.\  (Actually \ use \\
&  \sum\limits_{k=1}^{K-1}  \min{(\lvert \lvert F(k) - F(k-1) \rvert \rvert ,\lvert \lvert F(k+1)-F(k) \rvert \rvert)} )
\end{aligned}
\end{equation}

Recently, this list of features was expanded to 11, with two motion features (one taking the min above, the other not), DLM (called ADM) and four additional DLM features at the same scales (0-3) as the four scaled VIF features. To this list, FastVDO adds one or two more motion features, described below. Note that the Netflix motion features, computed only on the original video, have no information about the loss of motion fidelity in the compressed or processed video. Nevertheless, as it does carry motion information, it performs well in predicting visual quality when fully trained,  on test data such as the Netflix dataset \cite{b4-nflx}. Further enhancements of VMAF are reported in \cite{b10-stvmaf}, which improves temporal information, but adds complexity.  But in our tests, where we train VMAF from scratch, it performs well below the 90\% level (see Fig. 6), with the BVIHD dataset \cite{b14-bvihd} proving to be especially challenging (~76\% in both PCC and SRCC). It proves challenging for our variants as well.

\begin{figure}[htbp]
\centerline{\includegraphics[scale=0.40]{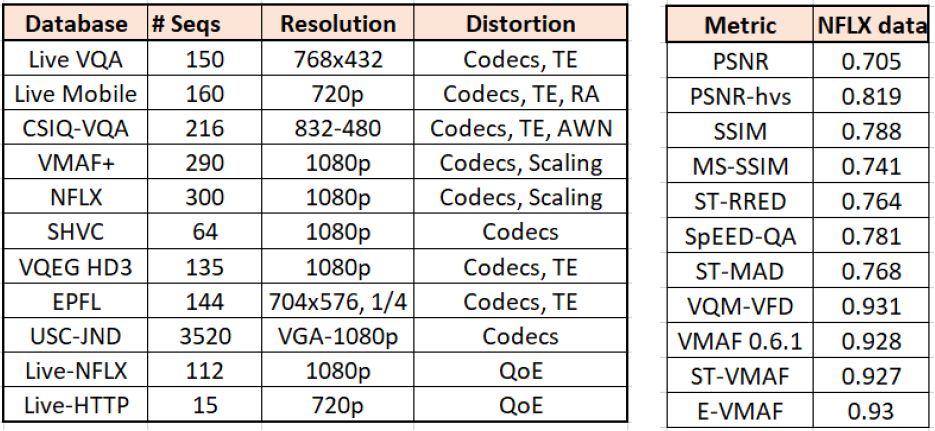}}
\caption{Some results from \cite{b10-stvmaf}, indicating VMAF and VQM-VFD perform well on Netflix databases}
\label{fig1}
\end{figure}

Relative to VMAF, we aim to improve on both the quality of motion representation, as well as the learning-based regressor engine. Specifically, for original video frames F(k), k=0,…,K, and processed (distorted) video frames G(k), k=0,…,K, since the temporal frame difference precisely corresponds to motion (all changes to pixels), we develop temporal motion based metrics using the difference of frame differences (equation 5). Let’s call this feature DM for differential motion. We retain the VIF and DLM features too. Like VMAF, we can use a range of features, from the base six (DM, DLM, VIF0-VIF3), up to a max of 13, where we add two variants of DM to the 11 current features of VMAF. We call our version of these features the FVMAF features.  Importantly, we greatly advance the learning-based regressor engine to include feedforward neural networks (NNs), besides the SVR.

\begin{equation}
\begin{aligned}
& DM = \lvert \lvert (F(k)-F(k-1))-(G(k)-G(k-1)) \rvert \rvert, \\
& the \ L1 \ norm \ (but \ can \ be \ L2, \ Lp, \ Entropy, \ etc).  
\end{aligned}
\end{equation}

\section{No Reference (NR) VQAs}

Historically, FR VQAs have been mainly image quality assessments (IQAs), applied per frame and averaged. PSNR and SSIM are common examples. This can also be done in the NR context, e.g., NIQE IQA \cite{b11-niqe}. The first completely blind NR VQA was released in 2016: Video Intrinsic Integrity and Distortion Evaluation Oracle (VIIDEO) \cite{b12-viid}, based on NIQE IQA. This is an explicit, purely algorithmic approach without any prior training. It relies on a theorized statistical feature of natural images \cite{b13-ruder}, captured as Natural Scene Statistics (NSS), and measured in the temporal domain. Analyzing frame differences, in local patches, they normalized the patch pixels by subtracting the mean, dividing by the standard deviation, modeling the resulting patch pixel data as a generalized Gaussian distribution, and estimating its shape parameters. Natural (undistorted) images would have Gaussian statistics, while the distortions would alter that shape parameter, which then leads to a distortion measure. An important fact is that this NR VQA already beats the common MSE, a FR VQA. Fig. 4 gives example images from FR datasets BVIHD \cite{b14-bvihd}, NFLX-2 \cite{b4-nflx}, and the NR dataset YouTube UGC \cite{b15-ytugc}. Note that the BVIHD dataset is challenging for VMAF; see Fig. 5. A follow-on NR VQA called SLEEQ \cite{b16-sleeq} was developed in 2018, which took the NSS concept further. First, as there is no original reference to compare to, they create a “self-referenced” comparator by blurring the compressed (or processed) video by a Gaussian blur, whose standard deviation then becomes a design parameter. The compressed and blurred compressed videos are then compared in patches, in both spatial and temporal domains, to create individual spatio-temporal distortion measures, which are then combined. As the recent paper \cite{b17-subugc} indicates, the performance of leading NR VQAs shows significant challenges; see Fig. 4.

Given these challenges, a comprehensive analysis of proposed methods was undertaken in 2020 \cite{b18-videval} for the NR case. In it, the authors reviewed a vast literature of NR algorithms, which they now viewed as merely providing features to process. They begin with no less than 763 features, then downselected to 60 features using learning methods of support vector machines (SVM) and Random Forests. These 60 features were then aggregated using a highly optimized support vector regressor (SVR) with hyperparameters optimization to achieve state-of-the-art performance in NR VQA \cite{b18-videval}. While impressive, its complexity is very high (though recently reduced in \cite{b25-videvallight}), something both this paper and RAPIQUE \cite{b20-rapique} aim to address. RAPIQUE uses a mix of fixed-function and neural net based features, creating a huge 3884-dimensional feature vector, yet offers some speed gains over VIDEVAL due to the nested structure of features.  Somewhat similarly, CNN-TLVQM \cite{b26-cnntlvqm} also uses a mix of fixed and CNN-based features for the NR case, and reports strong results. Meanwhile, we remark that \cite{b24-regclas} asks whether VQA should even be a regression problem, rather than a classification one, offering benchmarks for both. 

\section{FVQ: Towards a Unified VQA}

Combining insights from both the FR and NR cases, as exemplified by VMAF and SLEEQ, we arrive at a partial synthesis; see Fig. 2. In FR, we compare a processed video to the original; in NR, we compare it to a blurred processed video. In both cases, we input two videos, extract spatio-temporal features, pass them to a regressor engine, and obtain a quality score. The feature extractor and the regressor can both use learning-based methods such as neural nets. For complexity reasons we prefer fixed-function feature extractors for now, but note that CNN features are quite popular \cite{b21-vsfa}, \cite{b22-mdtvsfa}, \cite{b20-rapique}, and we will use them. Note that if desired, we can always add the output score of any NR algorithm as an additional feature in any NR or FR VQA. 

\subsection{FR, NR Feature Selection}
For high-quality FR databases, our FVMAF features are nearly the same as for VMAF (with a new motion feature DM); we can apply these same (type of) features in NR testing as well. But in the challenging NR VQA case for user generated content (UGC), our FVMAF features prove to be inadequate. So we utilize the powerful features from \cite{b18-videval} and \cite{b20-rapique}; see Figs. 5, 8, 9. We select a subset of features using a simple selection criterion, described here. We order the features according to a novel combined correlation coefficient measure S(f) (equation 6), and often select a subset of the top ranked features. For example, out of the 3884 RAPIQUE features, we may select say 200; see fig. 9. Here \textalpha \ = 0.5 by default.

\begin{equation}
\begin{aligned}
& S(f) = \alpha * \lvert PCC \rvert + (1-\alpha) * \lvert SRCC \rvert;\ \  \alpha = 0.5
\end{aligned}
\end{equation}

\subsection{Regression}
For regression, we use both SVRs and simple feedforward fully-connected NNs. For the SVR, we conduct a limited hyperparameter search, for parameters known as C, gamma, epsilon. For the neural net, we use a very simple fully connected feedforward network; as an example, with 13 features, we use a 6-80-64-1  network, with Relu activation, RMSProp optimizer, and Tensorflow 2.4.1, to aggregate the features (we occasionally use sigmoid activation in the last layer). Post feature extraction, an example {\bf total train/test simulation time} for 1k sims of  SVR takes only 10s to run (no-GPU) in FR, while 50 sims of NN took 127s on a laptop (i7-10750, 16GB RAM, RTX 2070 GPU). In NR, our SVR inference time with our reduced parameter search takes only 4s; post parameter search, only 0.01s.

For direct comparisons in trainability, we use the same train and test regimen as our comparators, and train from scratch. In the FR case, when comparing with VMAF, we use essentially the same features and SVR settings as VMAF; but change the motion feature, and improve some hyperparameters. Our fixed SVR parameters are (C, eps, gam) = (1000, 1, 0.1), while for moderate search, we search (C, eps, gam) in range ([10,100,900,1000],[.01,.05,.1,.5,1],[.0001,.001,.01,.1,1]). In the NR case, we use the 60 VIDEVAL features, the same SVR framework and hyperparameter set, or a no parameter neural network. We also test with subsets of say 100 of the RAPIQUE features (out of 3884).  Our results in the FR case indicate that both our SVR and NN methods outperform VMAF (fig. 6). And even when ignoring the reference on FR data, we get useful results with the same type of features and regressors, partially validating our unified approach (fig. 7). In the NR case with challenging UGC datasets, even when we compare to the fully trained VIDEVAL or RAPIQUE algorithms, our correlation scores are competitive, while greatly reducing hyperparameter search complexity during training (Figs. 8, 9). In fact, with both SVR and short NN, since today’s TPUs can process FFNNs in real-time on at least 1080p30 resolution, the main complexity in execution now lies in the feature extraction phase. Thus, greatly limiting the expensive feature extraction part is then critical to live usability.  For FR we mainly use just 6 features; for NR, we test with 10-400 features, drawn from VidEval, Rapique, or both.

\subsection{Results and Discussion}
Fig. 6 comprises our main results for the FR case. In FR, we use features identical to the VMAF features but with an improved motion feature, and use improved regression using an SVR, or a neural network. We obtain results across several datasets that exceed VMAF in both PCC and SRCC by roughly 5\%, and up to 15\%, which is a substantial gain. Achieving mostly over 90\% across datasets for both PCC and SRCC, with no prior training, this technology appears to be maturing. (But one dataset, BVIHD, proves to be challenging for both VMAF and our methods.) Gains are achieved with no increase in training/testing complexity with either the SVR with fixed parameters, or the NN with no parameter search. With moderate search using the SVR, we can obtain some further gains; see fig. 6.  Moreover, even if we ignore the reference videos and view these datasets as NR, we still obtain useful results, again using a NN regressor. It is this finding that partially validates our attempted synthesis of FR and NR in one framework. But this is still NR in the high-quality domain of professional video. 

In the true NR case of UGC data, we our FVMAF features are inadequate, and we leverage the impressive work in VIDEVAL and RAPIQUE in developing powerful features. But we work to reduce the feature sets, while still obtaining results close to these SOTA algorithms. Furthermore, we elucidate the contribution of individual features using a novel saliency measure.mWe note that we focus here mainly on compression and scaling loss, leaving aside generic transmission errors for now as they are less relevant.

\subsection{Focus on the No Reference Case}
Good progress has been made on the FR case, but much work remains for the growing NR case due to its challenges. Computationally, we note that the FR VMAF features are few, fixed-function, integerized, multi-threaded, and fast; the VIDEVAL features are not, nor are the RAPIQUE features, which while many, are somewhat faster. But both algorithms are currently only at the research level, in Matlab (as is CNN-TLVQM). However, recently there has been a significant speedup of VIDEVAL (VIDEVAL\_light \cite{b25-videvallight}) by downsampling feature extraction in space and time, with marginal loss.  Hard work remains in the NR case to achieve both the performance and execution speeds needed in live applications, but we make useful progress on that front, similar in spirit to RAPIQUE  \cite{b20-rapique}. 

Meanwhile, we now focus on what performance can be achieved using the powerful features from VIDEVAL and RAPIQUE, but also our novel feature selection method prior to regression, if we choose to eliminate hyperparameter search during training. We see from figure 9 that for the AllCombined NR dataset with 3165 videos, only a few of the most salient features provide most of the predictive performance, via SVR or NN. Two advantages of the NN method are that (a) it is fast on a modern GPU, and (b) it doesn't require hyperparameter search during training, significantly enhancing the potential for live application. We see from figures 8 and 9 that, either using the 60 VidEval features, or a 100-200 mixed VidEval and Rapique features, we can get to within 3\% of these algorithms on the AllCombined dataset, using a fixed architecture neural network, without any parameter search or parametric curve-fitted prediction. Our best NR result for the AllCombined dataset is with using all 60 VidEval features, and 120 top Rapique features (total of 180), obtaining PCC/SRCC of 0.766/0.764; see fig. 8. Moreover, the 50 cycles of training/testing in our neural network ran at least 20X faster than the SVR in a Google Colab tensorflow simulation environment. And our saliency analysis helps elucidate which features are the most informative. To achieve new state-of-the-art performance in NR going forward, our plan is to incorporate additional powerful features (such as from \cite{b26-cnntlvqm}), use more extensive parametric search and use curve-fitting for SVR prediction (as in VIDEVAL and RAPIQUE), or use the power of NN regression more fully.

\begin{figure}[htbp]
\centerline{\includegraphics[scale=0.28]{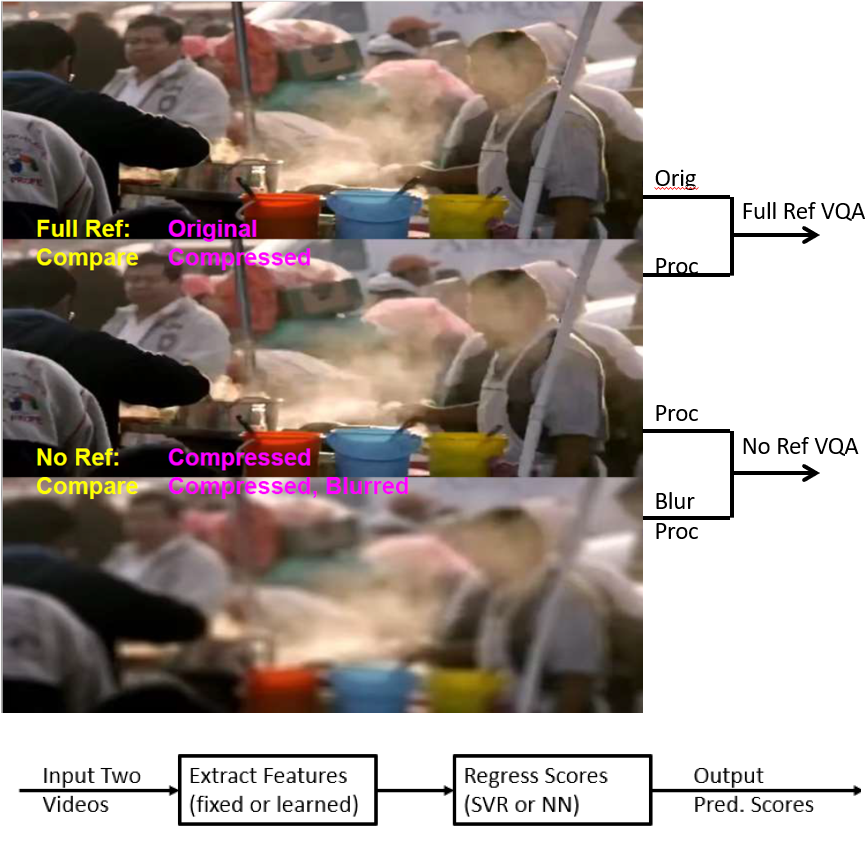}}
\caption{Outline of FastVDO Quality (FVQ) calculation. Two videos are input, either original and processed (FR case), or processed and blurred-processed (NR) case; features are extracted, which may be based on learning methods; and predicted scores are regressed, using a learning method such as SVR or a neural network (NN).}
\label{fig}
\end{figure}

\begin{figure}[htbp]
\centerline{\includegraphics[scale=0.40]{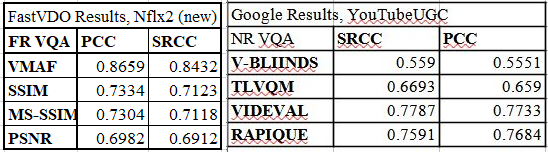}}
\caption{Example performance of known VQAs in (a) the FR case by FastVDO, no pretrained models; (b) the NR case, by UT-Google \cite{b20-rapique}, heavily trained models. While the FR case achieves useful levels of correlation with viewer ratings, the state of NR is  more challenging. Recent advances have made progress. Our FVQ achieves 0.847 in SRCC without pretraining; see Fig. 6.}
\label{fig}
\end{figure}

\begin{figure*}
\centerline{\includegraphics[scale=0.14]{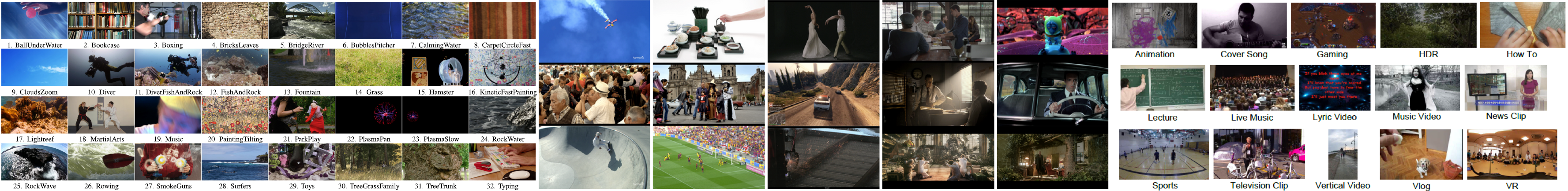}}
\caption{Example images from three databases under test: (a) BVI-HD \cite{b14-bvihd}; (b) NFLX-II \cite{b4-nflx}; and (c) YouTube UGC \cite{b15-ytugc}. The BVI-HD and NFLX-II databases have originals of high-quality, stable videos. The YouTube UGC data has mostly modest quality, user-generated videos, though it also has 4K HDR. }
\label{fig}
\end{figure*}

\begin{figure*}
\centerline{\includegraphics[scale=0.12]{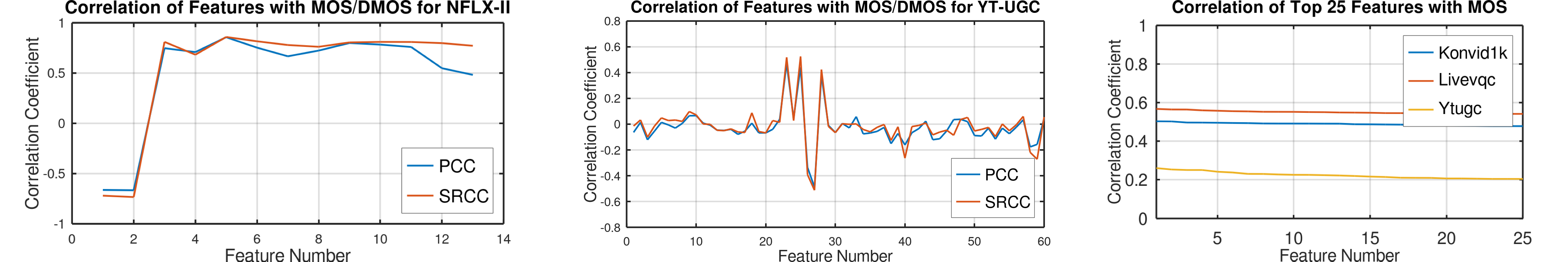}}
\caption{Correlation coefficients PCC and SRCC of VMAF features to user ratings: (a) to the NFLX-II data; (b) VidEval features on the YouTube UGC data; and (c) the mean of PCC and SRCC for top 25 features from the Rapique feature set (from 3884 features).  All VMAF features appear useful, and are used (fig. 6); in fig. 9, we use the 60 VidEval features.}
\label{fig}
\end{figure*}

\begin{figure*}
\centerline{\includegraphics[scale=0.46]{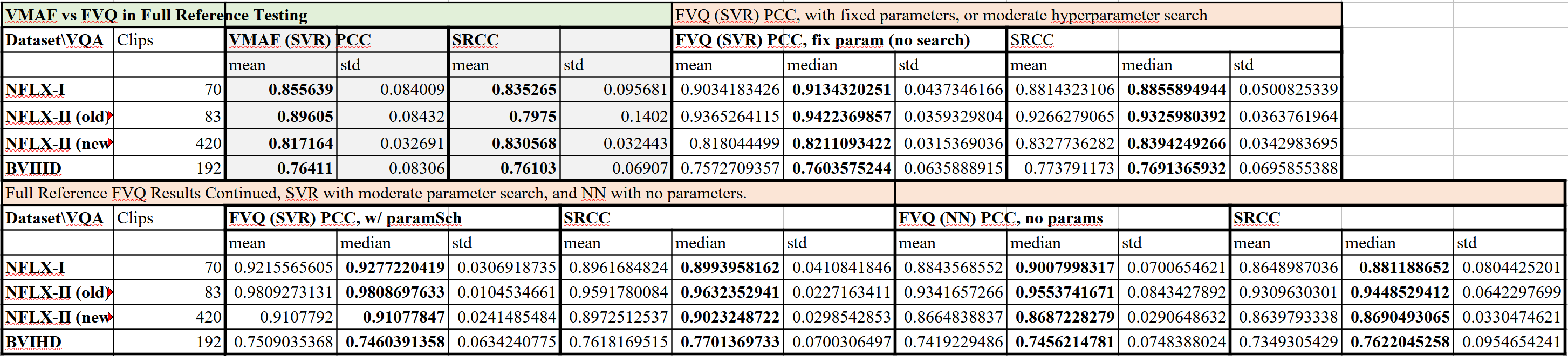}}
\caption{Simulation results for FVQ the full-reference (FR) testing, comparing to the well-known VMAF algorithm. In our tests, for a fair comparison, we train/test on each dataset, splitting the data randomly 80/20 for SVR, 85/15 for NN, and repeating the test 50 times. Our results show that we can outperform VMAF on these test datasets with both SVR and NN. We believe the gains come from both a better motion feature than VMAF, as well as better regression. While VMAF also regresses using an SVR, we obtain gains by using improved hyparameters (no search), as well as with moderate search. Using a modern GPU, our computationally fastest regressor engine is a NN (e.g., a simple 6-80-64-1 fully connected feedforward network, with Relu activation, and RMSProp optimization), which achieves excellent results.}
\label{fig}
\end{figure*}

\begin{figure}
\centerline{\includegraphics[scale=0.40]{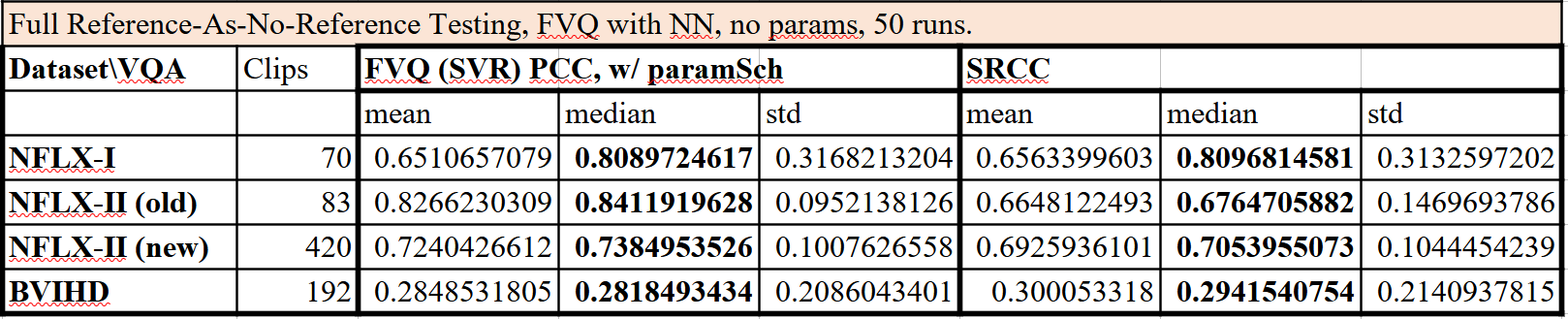}}
\caption{Simulations in the full reference databases, but ignoring the reference videos (full-as-no-ref). It is encouraging to see that even by working without the reference, but otherwise using the same framework (including the same feature types and regressor engines), we can obtain quite useful results. This partly validates our attempts at a unified framework for VQA. We remark that this approach may be applied even if the reference is not at hand, and can simplify the workflow more generally. We note that this application is still in the context of the high-quality, FR data. By contrast, it is much harder to obtain good performance in the true no reference case.}
\label{fig}
\end{figure}

\begin{figure*}[htbp]
\centerline{\includegraphics[scale=0.47]{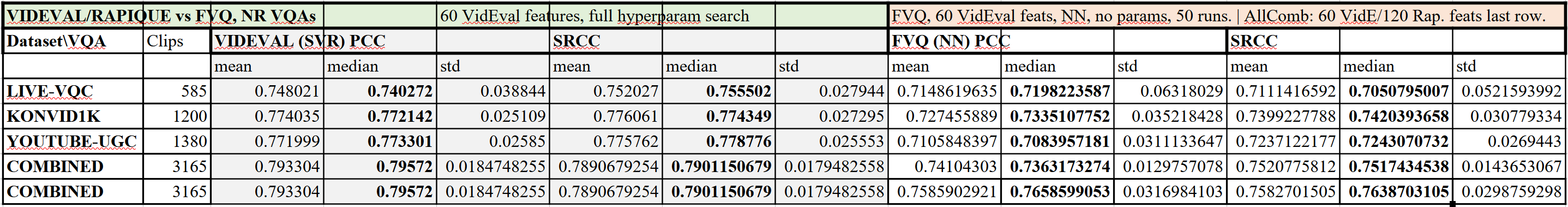}}
\caption{Simulation results in no reference testing, using several large no reference user generated content databases. Comparisons are to a state-of-the-art VIDEVAL algorithm, where we use the same 60 features, but a different regressor (a fixed NN architecture) and where we reduce complexity by avoiding any hyperparameter search. For the last row, we use 60 VidEval features, and 120 Rapique features (total of 180) to obtain our best results with the NN regressor, and no parameter search. The main value of our approach is to expedite training processing in modern GPU-enabled compute architectures. In our tests, the combined training and testing for 50 runs at least 20X faster than the SVR with full parameter search in Google Colab tensorflow computation. We also hope to be able to reach state-of-the-art results in the near future, by combining additional powerful and salient features in the regression.}
\label{fig}
\end{figure*}

\begin{figure*}[htbp]
\centerline{\includegraphics[scale=0.06]{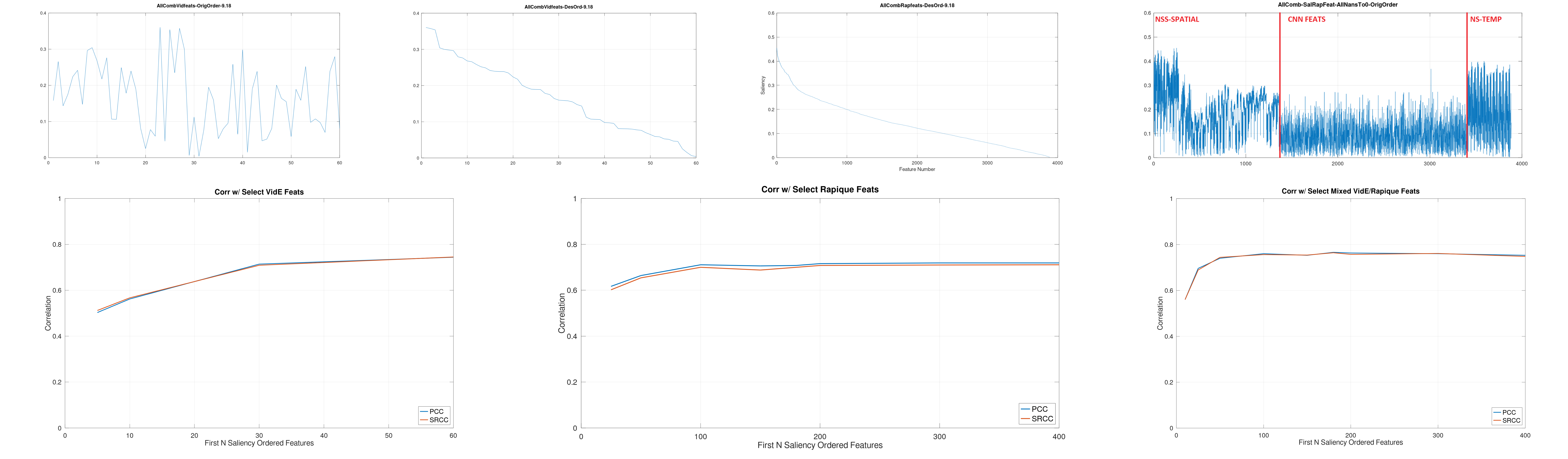}}
\caption{Top Row: Saliency S(f) of features from VidEval for the AllCombined NR dataset, in (a) original order, (b) descending Saliency order; Saliency of Rapique features for the AllCombined NR dataset, in (c) descending order, and (d) original order. Bottom Row: Prediction performance under a FastVDO neural network regressor, over 50 runs, using the first N Saliency-ordered features, from (a) VidEval, (b) Rapique, and (c) Mixed VidEval and Rapique. Our best results are with all 60 VidEval and the top 120 Rapique features, reported in the last row of fig. 8. While our approach does not achieve state-of-the-art performance, we do clarify that just a few of the most salient features from each algorithm are providing most of the predictive performance.}
\label{fig}
\end{figure*}

\begin{figure}[htbp]
\centerline{\includegraphics[scale=0.20]{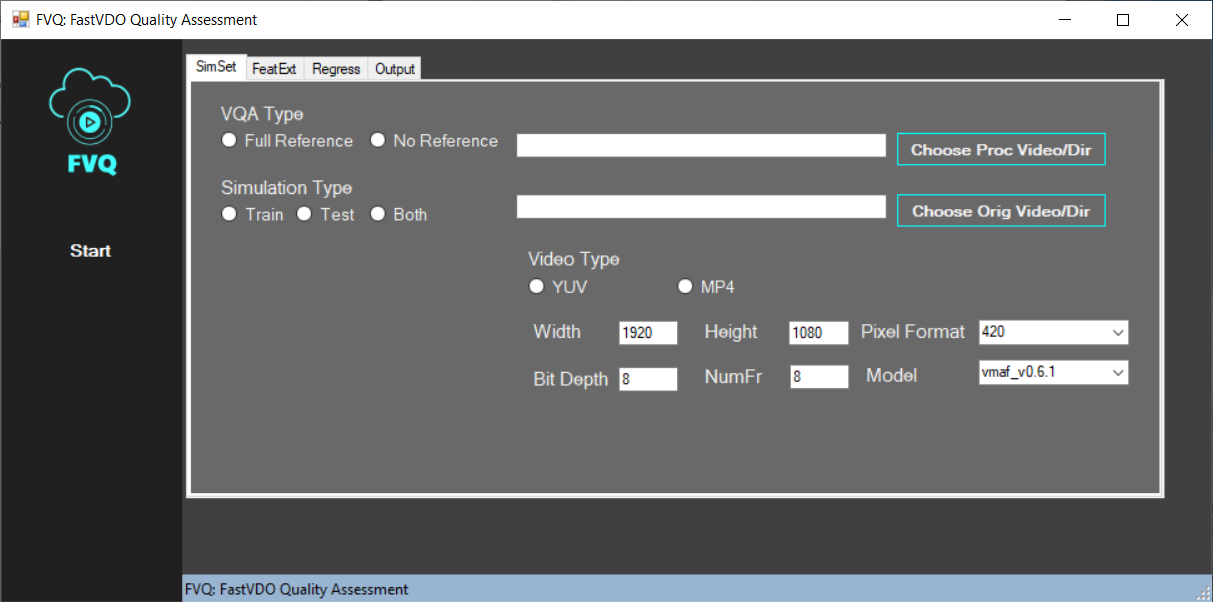}}
\caption{Screenshot of the FVQ application, capable of training and testing (or both, with random split), and computing VMAF as well as FVSVR and FVNN VQA scores, with parameter selection.}
\label{fig}
\end{figure}

\section{Conclusions}

We investigated an approach to assessing video quality in both FR and NR cases using a single framework, consisting of taking two input videos (original, processed; or processed, blurred-processed), evaluating a variety of (fixed) algorithmic features on these videos, and regressing these using an SVR or a feedforward NN to obtain a score. In the FR case, we are able to show improved trainability versus VMAF on example databases, using nearly the same feature set, with improvements in the motion feature and an SVR or a NN regressor. In the NR case, we reduce training complexity by eliminating hyperparameter search, yet achieve performance close to VidEval and Rapique on the AllCombined dataset. Further research remains to achieve a new state-of-the-art performance.

\end{document}